\documentclass[pra,twocolumn,showpacs,preprintnumbers,amsmath,amssymb,aps]{revtex4}
\usepackage{color}
\usepackage{bm, graphicx, amsmath}

\begin{document}
\title{A quantum controlled quantum eraser}
\author{J\'anos A. Bergou and Mark Hillery}
\affiliation{Department of Physics and Astronomy, Hunter College of CUNY, 695 Park Avenue, New York, NY 10065 \\Physics Program, Graduate Center of the City University of New York, 365 Fifth Avenue, New York, NY 10016}

\begin{abstract}
We show that the protocol known as quantum state separation can be used to transfer information between the phase and path of a particle in an interferometer.  When applied to a quantum eraser, this allows us to erase some, but not all, of the path information.  We can control how much path information we wish to erase.
\end{abstract}

\maketitle

\section{Introduction}
One of us (MH) joined the group of Marlan Scully in 1980, which was the same year Wolfgang became associated with the group. The other one (JB) became associated with group in 1982 and had the privilege of sharing an office with a young PhD student - Wolfgang, of course - in the famous "Theoretiker Container", a metallic container where all the theorist at the Max-Planck Institute for Quantum Optics were roasted during summers and frozen stiff during winters. Around this time, the hot topic in the group was the quantum eraser \cite{scully,hillery}.  In the papers that resulted from those discussions, it was shown how to erase the path information that had been acquired about a photon, which had scattered off of two atoms, and thereby restore the photon's interference pattern.  The quantum eraser as a sorting problem of measurement data has been discussed in \cite{englert} where a family of inequalities has also been derived relating quantitative measures of which way information and coherence. On the occasion of Wolfgang's 60th birthday, it seems appropriate for us to return to the quantum eraser.  What we want to show here is how to use the procedure of quantum state separation, originally developed by Chefles and Barnett \cite{chefles}, to move information between path information and the visibility of an interfering particle.

Consider a particle going through an interferometer, which can take one of two paths.  The paths correspond to the two orthogonal states $|1\rangle$ and $|2\rangle$.  With each path, we associate a detector state $|\eta_{j}\rangle$ for $j=1,2$.  The state $|\eta_{1}\rangle$ is the state of the path detectors if the particle is in path $1$, and similarly for $|\eta_{2}\rangle$.  These states are not assumed to be orthogonal, and we shall set $\langle\eta_{1}|\eta_{2}\rangle = s$, where $s$ is real and $0 \leq s \leq 1$.  The overlap, $s$, encodes how much path information we have.  If $s=0$ we have complete path information and if $s=1$ we have none.  The overall state of the particle and detectors is
\begin{equation}
\label{interstate}
|\psi\rangle = \frac{1}{\sqrt{2}}(|1\rangle |\eta_{1}\rangle + |2\rangle |\eta_{2}\rangle ).
\end{equation}
If the particle in path $1$ goes through a phase shifter, the state becomes
\begin{equation}
|\psi^{\prime}\rangle = \frac{1}{\sqrt{2}}( e^{i\phi}|1\rangle |\eta_{1}\rangle + |2\rangle |\eta_{2}\rangle ).
\end{equation}
and an interference pattern emerges at the output if we measure $X=|1\rangle\langle 2| + |2\rangle\langle 1|$ yielding
\begin{equation}
\langle \psi^{\prime}|X|\psi^{\prime}\rangle = s\cos\phi .
\end{equation}
The visibility of this pattern is just $s$.

By changing $s$, we can shift the balance between path information and visibility.  If we increase $s$ we increase the visibility and decrease the path information, and if we decrease it, we decrease the visibility and increase the path information.  In the original quantum eraser, $s$ was initially zero and it was subsequently increased to one.  Here we would like to show how it can be changed by an arbitrary amount.

\section{State separation and state compression} 
We will give a simple derivation of a quantum protocol that can probabilistically change the overlap of quantum states.  The original protocol due to Chefles and Barnett was called state separation because the interest was in reducing the overlap of the states \cite{chefles}.  They employed the general POVM formalism, and gave the solution for equal priors. It was later extended to a symmetric set of coherent states in \cite{dunjko1,dunjko2} and recently a complete solution for arbitrary priors was presented in parametric form \cite{bagan1}. In this later work the Neumark extension of the original POVM method \cite{neumark} was employed, which produces the same results in a more direct way, and we will use this formalism throughout this paper.  

Suppose we have two states, $|\psi_{1}\rangle$ and $|\psi_{2}\rangle$, where $\langle\psi_{1}|\psi_{2}\rangle = s$ and, with no loss of generality, we assume, as before, that $s$ is real and that $0 \leq s \leq 1$.  We want to transform these to two states, $|\phi_{1}\rangle$ and $|\phi_{2}\rangle$, called the signal states. We introduce the overlap of the signal states as $\langle\phi_{1}|\phi_{2}\rangle = t$, with $t$ real, and if $s \leq t \leq 1$ we call the protocol state compression and if $-1 \leq t \leq s$ we call the protocol state separation.  This can be accomplished by appending an ancilla and performing the unitary transformation on the extended system
\begin{equation}
U(|\psi_{j}\rangle_{a}|0\rangle_{b})=\sqrt{p}|\phi_{j}\rangle_{a}|0\rangle_{b} + \sqrt{1-p}|\gamma_{j}\rangle_{a}|1\rangle_{b} ,
\end{equation}
for $j=1,2$.  We assume that $0 \leq p \leq 1$ and $\langle\gamma_{1}|\gamma_{2}\rangle = r$ is real.  The states  $|\gamma_{1}\rangle$ and $|\gamma_{2}\rangle$ are called the idler states. 

To accomplish the transformation, we measure the ancilla, and if we obtain $|0\rangle_{b}$, then we have succeeded.  This occurs with probability $p$.  In order to relate all of the parameters, we take the inner product of the above equation with $j=1$ with the same equation with $j=2$ and find that
\begin{equation}
p=\frac{s-r}{t-r} .
\label{pC}
\end{equation}
The fact that $0 \leq p \leq 1$ implies that the numerator and denominator in the above expression have the same sign, they are both simultaneously positive or simultaneously negative. 

In the first case, both of them positive, we have $s \geq r$ and $t \geq r$ and, from the requirement $p \leq 1$, we also have $s \leq t$, so this case corresponds to state compression. In the second case, both of them negative, we have $s \leq r$ and $t \leq r$ and from the requirement $p \leq 1$, we also have $s \geq t$, so this case corresponds to state separation.  

\begin{figure}[ht]
      \centering
      \includegraphics[height=7 cm]{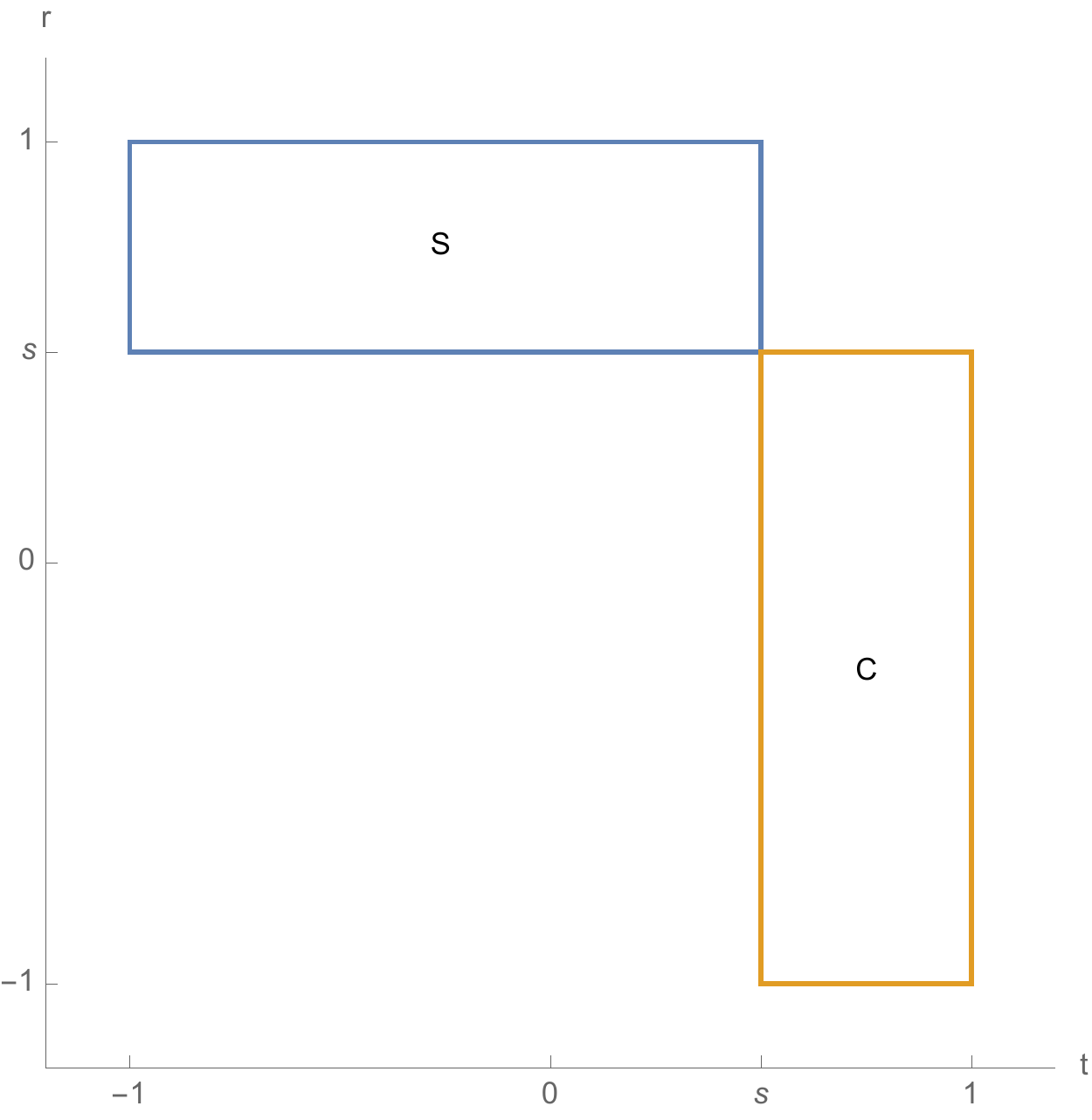}
      \caption{Physical regions of the overlap of the signal states, $t$, and the overlap of the idler states, $r$. In the blue region, denoted $S$, separation of the signal states occurs, $t < s$. As discussed after Eq. \eqref{S}, we only need to consider the part of the blue region that corresponds to $t \geq 0$. In the brown region, denoted $C$, compression of the signal states occurs, $t > s$. Note that the idler states exhibit the opposite behavior as the signal states. For the figure, the value of $s=0.5$ was used.}
      \label{Fig1}
\end{figure}

Summarizing, we have two distinct cases. If we increase the overlap between the states, which corresponds to state compression, we have the hierarchy
\begin{equation}
r \leq s \leq t
\end{equation} 
or, in more detail,
\begin{equation}
s \leq t \leq 1\ \ \ \ \ \mbox{and} \ \ \ \ \  -1 \leq r \leq s.
\label{C}
\end{equation}
If we decrease the overlap between the states, which corresponds to state separation, we have the hierarchy
\begin{equation}
t \leq s \leq r
\end{equation}
or, in more detail,
\begin{equation}
s \leq r \leq 1 \ \ \ \ \ \mbox{and} \ \ \ \ \  -1 \leq t \leq s.
\label{S}
\end{equation}
Equations \eqref{C} and \eqref{S} select the physically allowed regions of the $\{t,r\}$ parameter plane. They are depicted in Fig. \ref{Fig1}. For state separation, the actual range can be restricted to $0 \leq t \leq s$.   For $-s \leq t \leq 0$, we can accomplish the same amount of separation for $|t|$, since only the magnitude of the overlap matters.  If $-1 \leq t < -s$, the magnitude of the overlap is increasing rather than decreasing, which is the opposite of the desired result. Consequently, for state separation we can confine our attention to values $0 \leq t \leq s$.

In the following, we look at the two cases, increasing the overlap (state compression) and decreasing the overlap (state separation), in more detail.  

\subsection{State compression: Increasing the overlap, $t \geq s$ and $r \leq s$}

\begin{figure}[ht]
      \centering
      \includegraphics[height=3.9 cm]{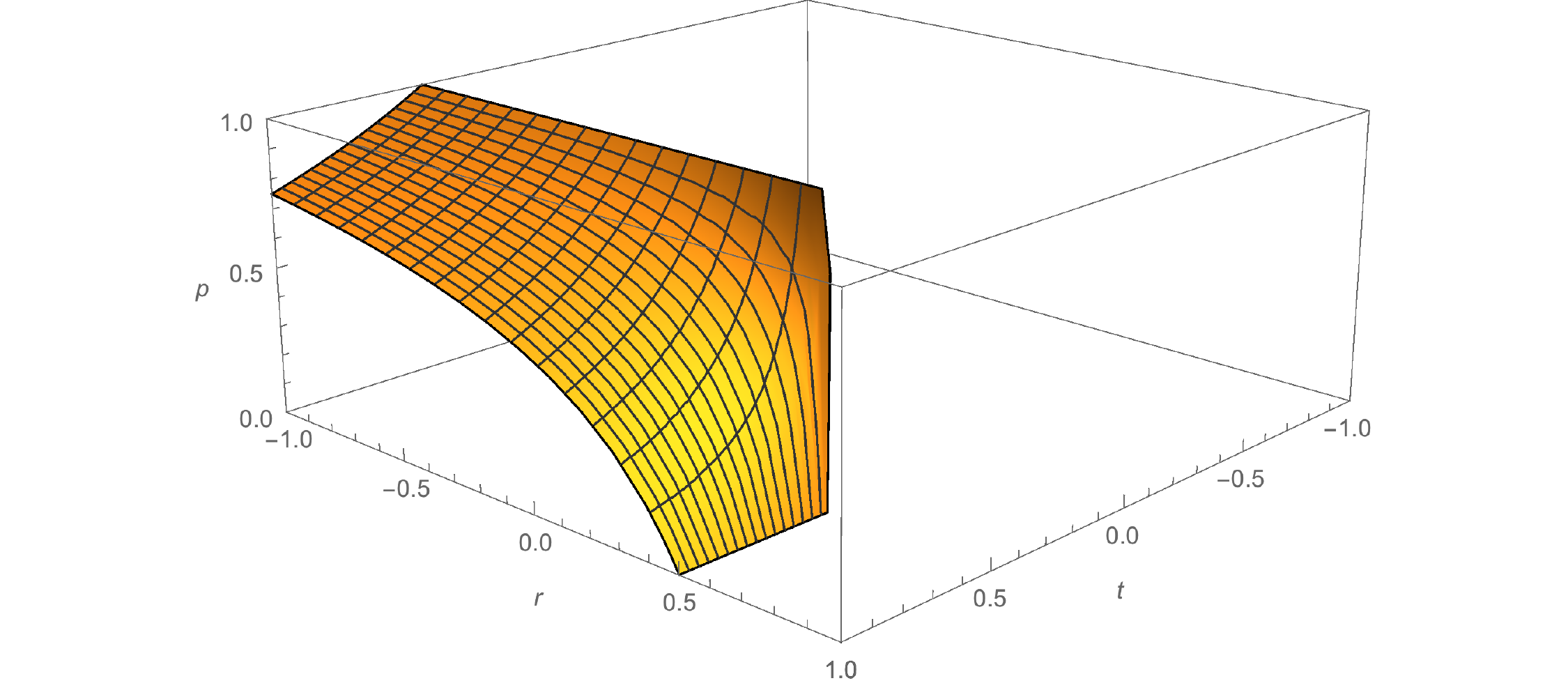}
      \caption{The probability of success for state compression, $p$, vs. the overlap of the signal states, $t$, and the overlap of the idler states, $r$. For the figure, the value of $s=0.5$ was used. The probability is monotonically increasing with decreasing $r$ and monotonically decreasing with increasing $t$. Note that $p=1$ along the $t=s$ line and $p=0$ along the $r=s$ line.}
      \label{Fig2}
\end{figure}

First, we discuss the case of state compression, corresponding to the brown region, $C$, in Fig. \ref{Fig1}, as defined by the inequalities in Eq. \eqref{C}. In this case, we can directly use Eq. \eqref{pC} for the success probability of state compression. Two features are obvious: $p=0$ along the $s=r$ segment, which is the upper edge of the brown region and $p=1$ along the $t=s$ segment which is the left edge of the brown region. The probability as a function of the overlap of the signal states, $t$, and the overlap of the idler states, $r$, is displayed in Fig \ref{Fig2}.

From the figure it is clear that, as a function of $r$, the success probability of state compression $p$ reaches its maximum when $r=-1$. To see this analytically, from Eq. \eqref{pC} we can express $p$ as
\begin{equation}
\label{pComp}
p=1 - \frac{t-s}{t-r} .
\end{equation}
In order to make $p$ as large as possible, we have to make $r$ as small as possible, so we choose $r=-1$, yielding
\begin{equation}
p=\frac{1+s}{1+t} =\frac{1+s}{1+s+\Delta} ,
\label{pmaxC}
\end{equation}
where $\Delta=t-s>0$ is the change in the overlap.  

Most remarkably, it is also the absolute value of the change in the visibility of the interference pattern, when the state compression protocol is applied to the states $|\eta_{1}\rangle$ and $|\eta_{2}\rangle$ in Eq. \eqref{interstate}. When we drive the path states closer, increasing their overlap and, consequently, decreasing their distinguishability, we enhance the visibility of the interference pattern.   

\subsection{State separation: decreasing the overlap, $t \leq s$ and $r \geq s$}

\begin{figure}[ht]
      \centering
      \includegraphics[height=4 cm]{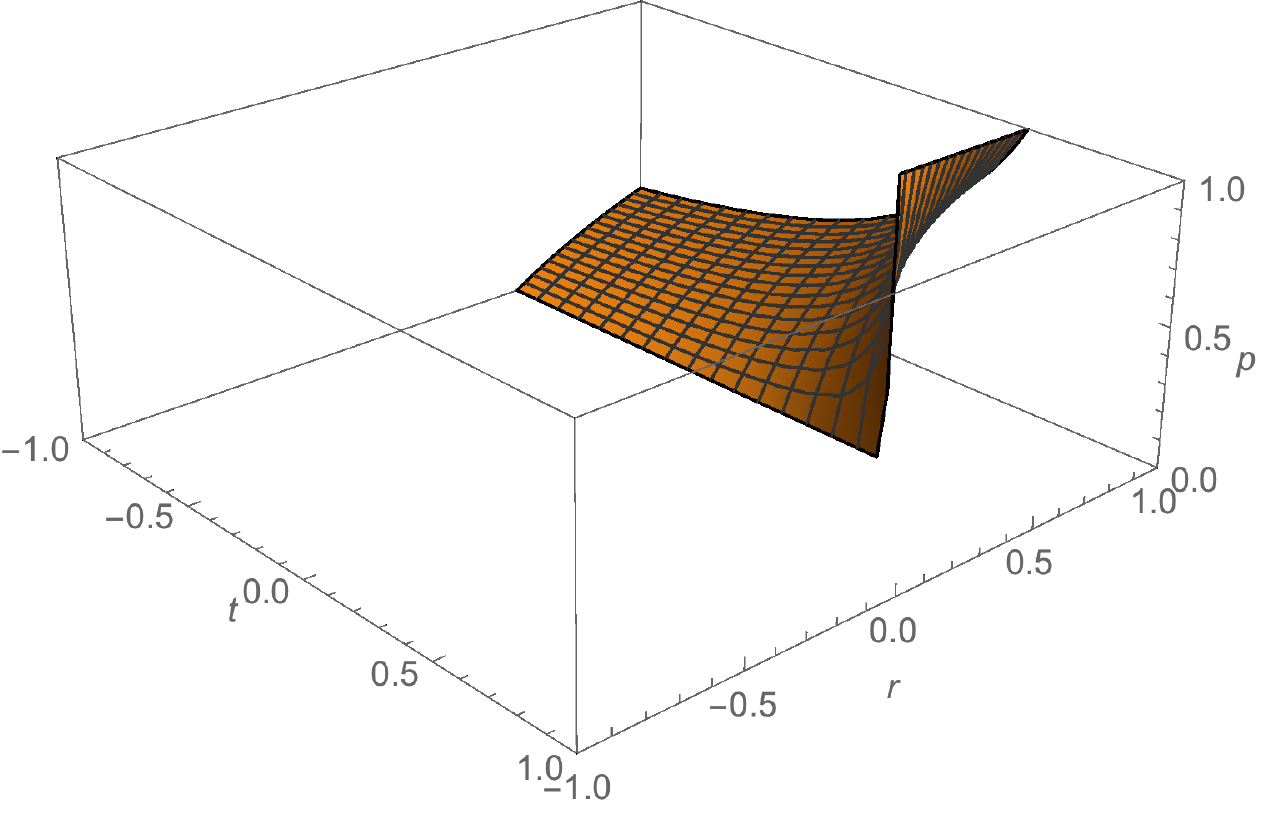}
      \caption{The probability of success for state separation, $p$, vs. the overlap of the signal states, $t$, and the overlap of the idler states, $r$. For the figure, the value of $s=0.5$ was used. The probability is monotonically increasing with increasing $r$ and monotonically decreasing with decreasing $t$. As discussed in connection with Fig. \ref{Fig1}, we only need to consider the $t \geq 0$ portion of the plot for optimal separation. Note that $p=1$ along the $t=s$ line and $p=0$ along the $r=s$ line.}
      \label{Fig3}
\end{figure}

Next, we discuss the case of state separation, corresponding to the blue region, $S$, in Fig. \ref{Fig1}, as defined by the inequalities in Eq. \eqref{S}. In this case, it is more convenient to use a version of Eq. \eqref{pC} for the success probability of state separation, where we multiply both the numerator and the denominator with $-1$. Two features are again immediately obvious: $p=0$ along the $s=r$ segment, which is the lower edge of the blue region and $p=1$ along the $t=s$ segment which is the right edge of the blue region. The probability as a function of the overlap of the signal states, $t$, and the overlap of the idler states, $r$, is displayed in Fig \ref{Fig3}.

From the figure it is clear that, as a function of $r$, the success probability of state separation $p$ reaches its maximum when $r=1$. To see this analytically, from Eq. \eqref{pC} we can express $p$ as
\begin{equation}
\label{pSep}
p=1 - \frac{s-t}{r-t} .
\end{equation}
In order to make $p$ as large as possible, we have to make $r$ as large as possible, so we choose $r=1$, yielding
\begin{equation}
p=\frac{1-s}{1-t} = \frac{1-s}{1+\Delta -s} ,
\label{pmaxS}
\end{equation}
where $\Delta=s-t>0$ is the change in the overlap.  

Most remarkably, the change in the visibility of the interference pattern is $-\Delta$, when the state separation procedure is applied to the states $|\eta_{1}\rangle$ and $|\eta_{2}\rangle$ in Eq. \eqref{interstate}. When we increase the separation of the path states, decreasing their overlap and, consequently, increasing their distinguishability, we reduce the visibility of the interference pattern.   

To close this section, we note that if the procedure fails, we end up with the idler states $|\gamma_{1}\rangle$ and $|\gamma_{2}\rangle$. In the case we were trying to increase the overlap, they satisfy $\langle\gamma_{1}|\gamma_{2}\rangle =-1$, if we use the optimum value, and in the case we were trying to decrease the overlap, they satisfy  $\langle\gamma_{1}|\gamma_{2}\rangle =1$, if we use the optimum value.  In the case we were trying to increase the overlap, and we fail, we certainly do increase the overlap, but for a state in an interferometer, we destroy all of the path information rather than just decrease it.

 \section{Connection to the quantum eraser}
 
We, now, have a method that will probabilistically change the amount of path information or, correspondingly, the visibility, of a particle inside an interferometer.  The original quantum eraser corresponded to the case $s=0$ and $\Delta = 1$, and the  formula given by Eqs. \eqref{pmaxC} and \eqref{pmaxS} tells us that this can be accomplished with a probability of $1/2$, which is in accord with the original result.  We can increase the probability of success by being less ambitious and erasing some, but not all, of the path information.  We can also do the reverse and increase the path information at the expense of reducing the visibility.  For example, if $s=3/4$ and $\Delta = 3/4$, we can obtain perfect path information with a probability of $1/4$.

The visibility can be interpreted as phase information \cite{coles,bagan2}.  Suppose one of the two states
\begin{equation}
|\psi_{\pm}\rangle = \frac{1}{\sqrt{2}}(|1\rangle_{a} |\eta_{1}\rangle_{b} \pm |2\rangle_{a} |\eta_{2}\rangle_{b} ),
\end{equation}
is split between two observers, Alice, who has access to the particle, and Bob, who has access to the detectors.  Alice's task is to determine which of the two states she has.  This means she has to discriminate between the two reduced density matrices.
\begin{equation}
\rho_{\pm}=\frac{1}{2}\left( \begin{array}{cc} 1 & \pm s \\ \pm s & 1 \end{array} \right) .
\end{equation}
If she uses minimum-error state discrimination to do this \cite{bergou}, her probability of succeeding is
\begin{equation}
P_{\phi}=\frac{1}{2}(1+s).
\end{equation}
Therefore, the visibility indicates how well Alice can discriminate between states with different relative phases between their two parts.  This means the procedure we have outlined can be viewed as moving information around between path information and phase information.

One might think that the state separation procedure could help Alice better discriminate the states $|\psi_{\pm}\rangle$.  Bob could apply the procedure to his detector states to increase $s$ and thereby increase Alice's success probability.  This, however, won't work.  By performing operations only in his laboratory and not communicating with Alice, Bob cannot change Alice's reduced density matrix, otherwise superluminal communication would be possible.  A short calculation of the effect of Bob applying the state separation procedure on his states shows, as expected, that Alice's reduced density matrix does not change.  If, however, the procedure does succeed, and Bob communicates this to Alice, then her chance of successfully discriminating the states does increase.

\section{A possible implementation}
One way of implementing the above procedure is to use photon polarization and a two-photon state, such as those produced by parametric down conversion \cite{gao}.  One photon is the particle going through the interferometer, and the other serves as the detector.  In particular, a state of the form
\begin{eqnarray}
|\psi\rangle & = & \frac{1}{\sqrt{2}} [a^{\dagger}_{H}(\cos\theta b^{\dagger}_{H} + \sin\theta b^{\dagger}_{V}) \nonumber \\ 
& & + a^{\dagger}_{V}(\cos\theta b^{\dagger}_{H} - \sin\theta b^{\dagger}_{V})] |0\rangle ,
\end{eqnarray}
could be used.  The operators $a^{\dagger}$ and $b^{\dagger}$ are creation operators for two modes, and each mode has two polarization states, H (horizontal) and V (vertical).  The state $|0\rangle$ is the vacuum.  The states $a^{\dagger}_{H}|0\rangle$ and $a^{\dagger}_{V}|0\rangle$ are analogous to the path states in Eq.\ (\ref{interstate}), and the $b$-mode states, 
\begin{equation}
|\psi_{b\pm}\rangle = (\cos\theta b^{\dagger}_{H} \pm \sin\theta b^{\dagger}_{V})|0\rangle ,
\end{equation}
are the analogs of the detector states, $|\eta_{1}\rangle$ and $|\eta_{2}\rangle$.  

\begin{figure}[ht]
\centering
\includegraphics[width=0.47\textwidth]{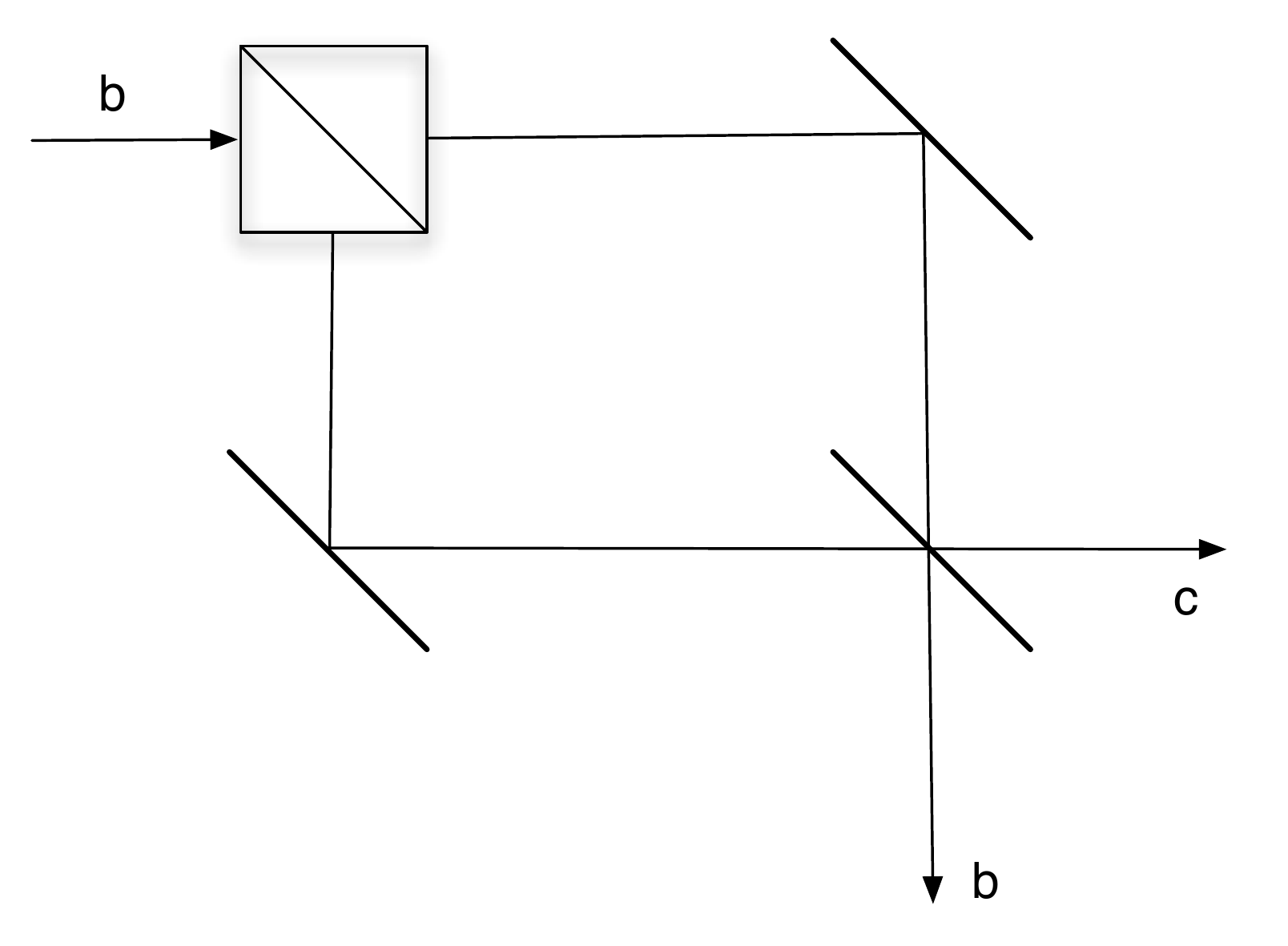} 
\caption{A network to implement state separation.  The photon in mode $b$ passes through a polarizing beam splitter and then a standard beam splitter.  We measure mode $c$ at the output, and if we detect no photon, the procedure has succeeded.}
\label{network}
\end{figure}

The issue now is how to construct an apparatus that will probabilistically change the overlap of $|\psi_{b+}\rangle$ and $|\psi_{b-}\rangle$. There are, of course, many possible physical implementations of this general quantum device. A particularly simple one, entirely in terms of a linear optical interferometer operating at the single-photon level and polarization optics, is depicted in Fig. \ref{network}.  

First send the $b$ mode into a polarizing beam splitter, which transmits the V polarization and reflects the H polarization.  This gives us, for the $|\psi_{b+}\rangle$ state (the analysis for the $|\psi_{b-}\rangle$ state is similar)
\begin{equation}
|\psi_{b+}\rangle \rightarrow (\cos\theta c^{\dagger}_{H} + \sin\theta b^{\dagger}_{V})|0\rangle ,
\end{equation}
where the $c$ mode is the reflected mode from the polarizing beam splitter.  The $b$ and $c$ modes are now combined at a standard beam splitter with transmission coefficient $\tilde{t}$ and reflection coefficient $\tilde{r}$, both of which we assume to be real.  The result is
 
\begin{eqnarray}
|\psi_{b+}\rangle & \rightarrow & (\tilde{r}\cos\theta b^{\dagger}_{H}+ \tilde{t}\sin\theta b^{\dagger}_{V})|0\rangle \nonumber \\
& & + (\tilde{t}\cos\theta c^{\dagger}_{H} - \tilde{r}\sin\theta c^{\dagger}_{V})|0\rangle.
\end{eqnarray}

We now measure the $c$ mode, and if no photon is detected, then the procedure has succeeded.  This procedure results in the state transformation
\begin{equation}
|\psi_{b\pm}\rangle \rightarrow \frac{1}{\sqrt{p}}(\tilde{r}\cos\theta b^{\dagger}_{H}\pm \tilde{t}\sin\theta b^{\dagger}_{V})|0\rangle ,
\end{equation}
where $p= (\tilde{r}\cos\theta )^{2} +  (\tilde{t}\sin\theta )^{2}$ is the probability of the procedure succeeding.

The initial overlap between the states $|\psi_{b\pm}\rangle$ is
\begin{equation}
s=\cos^{2}\theta - \sin^{2}\theta ,
\end{equation}
while the final overlap is
\begin{equation}
t=\frac{1}{p} ( \tilde{r}^{2} \cos^{2}\theta - \tilde{t}^{2}\sin^{2}\theta ) .
\end{equation}
The condition that $t<s$ implies that $|\tilde{r}|^{2} <1/2$, and conversely, $t>s$ implies that $|\tilde{r}|^{2}>1/2$.  Therefore, this setup can either decrease or increase the overlap between the states, and it can be used to shift information between paths and phases in a quantum eraser based on polarization. 

\section{Conclusion}
We have shown that the procedure of state separation can be used to move information about a particle in an interferometer between information about its path and information about its relative phase in the two branches, which is directly related to the visibility of the interference pattern produced by the particle.  The standard quantum eraser, in which the path information is eliminated, is an extreme example of this.  What we have shown here is that it is not necessary to erase all of the path information, one can erase some of it, or one can even increase it.

One of the things both of us share with Wolfgang is a lifelong (or at least since university) interest in quantum physics.  We hope this small example is a fitting present for Wolfgang on the occasion of his 60th birthday.


\begin{thebibliography}{99}
\bibitem{scully} M.~Scully and K.~Dr\"{u}hl, Phys.\ Rev.\ A {\bf 25}, 2208 (1982).
\bibitem{hillery}M. Hillery, M.O. Scully, On State Reduction and Observation in Quantum Optics: Wigner?s Friends and Their Amnesia, in Quantum Optics, Experimental Gravity, and Measurement Theory, NATO ASI Series, Vol. {\bf B94}, P. Meystre, M.O. Scully Eds. (Plenum Press, New York, 1983).
\bibitem{englert} B.-G. Englert and J. A. Bergou, Optics Comm. {\bf 179}, 337 (2000).
\bibitem{chefles} A.~Chefles and S.~Barnett, J.\ Phys.\ A {\bf 31}, 10097 (1998).
\bibitem{dunjko1}V. Dunjko and E. Andersson, J. Phys. A: Math. Theor. {\bf 45}, 365304 (2012).
\bibitem{dunjko2}V. Dunjko and E. Andersson, Phys. Rev. A {\bf 86}, 042322 (2012).
\bibitem{bagan1}E. Bagan, V. Yerokhin, A. Shehu, E. Feldman, and J. A. Bergou, New J. Phys. {\bf 17}, 123015 (2015).
\bibitem{neumark}M. Neumark, Izv. Akad. Nauk SSSR Ser. Mat. {\bf 4}, 53 (1940). For a more tutorial version see J. A. Bergou  and M. Hillery, {\emph{Introduction to the Theory of Quantum Information Processing}} (Graduate Texts in Physics) (Springer, New York,  2013).
\bibitem{coles} P.~Coles, J.~Kaniewski, and S.~Wehner, Nature Communications {\bf 5}, 5814 (2014).
\bibitem{bagan2} E.~Bagan, J.~Calsamiglia, J.~A.~Bergou, and M.~Hillery, Phys.\ Rev.\ Lett.\ {\bf 120}, 050402 (2018).
\bibitem{bergou} Discrimination of quantum states by J\'anos A. Bergou, Ulrike Herzog, and Mark 
Hillery in \emph{Quantum State Estimation} edited by M.\ G.\ A.\ Paris and J.\ \v{R}eha\v{c}ek
(Springer, Berlin, 2004).
\bibitem{gao}J. Gao, Zh-Q. Jiao, Ch-Q. Hu, L-F. Qiao, R-J. Ren, Zh-H. Ma, Sh-M. Fei, V. Vedral, and X-M. Jin, arxiv:1703.08026 (2017).

\end{thebibliography}
\end{document}